\documentclass[journal,10pt]{IEEEtran}
\usepackage{cite}
\ifCLASSINFOpdf
\else
   \usepackage[dvips]{graphicx}
   \DeclareGraphicsExtensions{.eps}
\fi
\usepackage[cmex10]{amsmath}
\usepackage{mathabx}
\begin{document}
%
% paper title
% Titles are generally capitalized except for words such as a, an, and, as,
% at, but, by, for, in, nor, of, on, or, the, to and up, which are usually
% not capitalized unless they are the first or last word of the title.
% Linebreaks \\ can be used within to get better formatting as desired.
% Do not put math or special symbols in the title.
\title{Pulse Shaping Methods for OQAM/OFDM and WCP-COQAM}
%
%
% author names and IEEE memberships
% note positions of commas and nonbreaking spaces ( ~ ) LaTeX will not break
% a structure at a ~ so this keeps an author's name from being broken across
% two lines.
% use \thanks{} to gain access to the first footnote area
% a separate \thanks must be used for each paragraph as LaTeX2e's \thanks
% was not built to handle multiple paragraphs
%

\author{Ali~Bulut~\"{U}\c{c}\"{u}nc\"{u},~\IEEEmembership{Student Member,~IEEE},
        Ali~\"{O}zg\"{u}r~Y{\i}lmaz,~\IEEEmembership{Member,~IEEE}% <-this % stops a space
\thanks{The authors are with the Department of Electrical and Electronics Engineering, Middle East Technical University, Ankara, Turkey (e-mail: ucuncu@metu.edu.tr, aoyilmaz@metu.edu.tr)}}

%\author{Author 1,
%		Author 2% <-this % stops a space
%\thanks{}}

% note the % following the last \IEEEmembership and also \thanks - 
% these prevent an unwanted space from occurring between the last author name
% and the end of the author line. i.e., if you had this:
% 
% \author{....lastname \thanks{...} \thanks{...} }
%                     ^------------^------------^----Do not want these spaces!
%
% a space would be appended to the last name and could cause every name on that
% line to be shifted left slightly. This is one of those "LaTeX things". For
% instance, "\textbf{A} \textbf{B}" will typeset as "A B" not "AB". To get
% "AB" then you have to do: "\textbf{A}\textbf{B}"
% \thanks is no different in this regard, so shield the last } of each \thanks
% that ends a line with a % and do not let a space in before the next \thanks.
% Spaces after \IEEEmembership other than the last one are OK (and needed) as
% you are supposed to have spaces between the names. For what it is worth,
% this is a minor point as most people would not even notice if the said evil
% space somehow managed to creep in.

% The paper headers
\markboth{IEEE Signal Processing Letters, Submitted Draft 2015}%
{Shell \MakeLowercase{\textit{et al.}}: Bare Demo of IEEEtran.cls for Journals}
% The only time the second header will appear is for the odd numbered pages
% after the title page when using the twoside option.
% 
% *** Note that you probably will NOT want to include the author's ***
% *** name in the headers of peer review papers.                   ***
% You can use \ifCLASSOPTIONpeerreview for conditional compilation here if
% you desire.

% If you want to put a publisher's ID mark on the page you can do it like
% this:
%\IEEEpubid{0000--0000/00\$00.00~\copyright~2014 IEEE}
% Remember, if you use this you must call \IEEEpubidadjcol in the second
% column for its text to clear the IEEEpubid mark.

% use for special paper notices
%\IEEEspecialpapernotice{(Invited Paper)}

% make the title area
\maketitle

% As a general rule, do not put math, special symbols or citations
% in the abstract or keywords.
\begin{abstract}
GFDM is a new modulation format whose advantages compared to OFDM reportedly make it a preferable modulation format for 5G. However, the non-orthogonal nature of GFDM with matched filtering (MF) receiver for pulses with good time-frequency localization is one of its disadvantages, leading to the proposal of WCP-COQAM, employing offset quadrature amplitude modulation (OQAM). In this paper, we prove that a pulse satisfying orthogonality conditions for OQAM-OFDM will also satisfy orthogonality with WCP-COQAM, thus the pulse design methods developed for OQAM-OFDM can also be used with WCP-COQAM. This statement is also verified by the simulation based results. \end{abstract}

% Note that keywords are not normally used for peerreview papers.

% For peer review papers, you can put extra information on the cover
% page as needed:
% \ifCLASSOPTIONpeerreview
% \begin{center} \bfseries EDICS Category: 3-BBND \end{center}
% \fi
%
% For peerreview papers, this IEEEtran command inserts a page break and
% creates the second title. It will be ignored for other modes.
\IEEEpeerreviewmaketitle

\section{Introduction}
% The very first letter is a 2 line initial drop letter followed
% by the rest of the first word in caps.
% 
% form to use if the first word consists of a single letter:
% \IEEEPARstart{A}{demo} file is ....
% 
% form to use if you need the single drop letter followed by
% normal text (unknown if ever used by IEEE):
% \IEEEPARstart{A}{}demo file is ....
% 
% Some journals put the first two words in caps:
% \IEEEPARstart{T}{his demo} file is ....
% 
% Here we have the typical use of a "T" for an initial drop letter
% and "HIS" in caps to complete the first word.
\IEEEPARstart{G}{eneralized} frequency division multiplexing (GFDM) is a recently proposed modulation technique which is amongst the candidate modulation formats to be used in future communication standards. This has many reasons. Firstly, it has been shown to possess lower out-of-band (OOB) radiation, compared to orthogonal frequency division multiplexing (OFDM)\cite{GFDM_5G}. The reason for high OOB radiation for OFDM compared to GFDM is that OFDM has rectangular pulse, which leads to large tails in the frequency domain, whereas pulse shapes other than the rectangular pulse can be utilized in GFDM. Pulses that have better frequency localization enable lower OOB radiation which may benefit cognitive radio based dynamic spectrum access solutions among others. Furthermore, the tail biting technique used in GFDM make the cyclic prefix length to be independent of the length of the transmit and receiver filters. This reduces the overhead caused by the cyclic prefix, which can be significant since short signal length is required for low latency applications. \cite{nekovee_low_latncy}. Another advantage of using GFDM is related to its flexible frame structure. By changing the number of time slots or subcarriers in a GFDM frame, it can cover both CP-OFDM or SC-FDE \cite{GFDM_5G}, depending on the requirements of the application. These merits of GFDM makes it a strong physical layer modulation technique candidate for future communication standards.

Despite its abovementioned advantages, GFDM has also some drawbacks. One of these is that for a GFDM scheme with matched filter (MF) receiver, if no inter-carrier interference (ICI) or inter-symbol interference (ISI) is desired, pulses such as rectangular or Dirichlet pulses should be used, which do not have good time-frequency localization. In fact, this results from Balian-Low theorem which states that there is no time-frequency well localized pulse when the lattice density is equal to 1 and this theorem cannot be circumvented if quadrature-amplitude modulation (QAM) is used as in GFDM. Therefore, zero-forcing (ZF), minimum mean square error (MMSE) or successive interference cancellation (SIC) receivers are employed in GFDM, to cancel ICI or ISI \cite{GFDM_ZF_MMSE}, \cite{GFDM_SIC}. However, such receivers also come with some drawbacks. Zero-forcing receiver suffers from noise enhancement, which will degrade the error rate performance. MMSE receiver aims a balance between noise and the interference. However, it does not achieve capacity without SIC. SIC receiver increases the complexity of GFDM receiver further. Therefore, in order to be able to employ pulses that have good localization in time and frequency with MF receiver, with no ISI or ICI, an extension of GFDM with offset-QAM (OQAM), is proposed as windowed cyclic prefix circular OQAM (WCP-COQAM) in \cite{lin2014multi}. It has also been called as TS-OQAM-GFDM in \cite{TS_OQAM_GFDM}. The reason behind using OQAM instead of QAM is to circumvent the Balian-Low theorem by lattice staggering. This approach is first used to synthesize orthogonal signals in \cite{Chang_OQAM},~\cite{Saltzberg_OQAM}. Until recently, such modulation schemes are referred to as OQAM/OFDM or staggered multitone (SMT). The proposal of WCP-COQAM is to achieve the aforementioned advantages of GFDM and at the same time to use pulses with good time frequency localization with MF receiver thanks to the use of OQAM. The advantage of using pulses with good time-frequency localization stems from their robustness against time-frequency dispersive channels or frequency synchronization errors such as carrier frequency offset (CFO) and their lower OOB emissions \cite{Bolkskei_TFL},\cite{haas1997time}.

Although in \cite{lin2014multi}, WCP-COQAM is described with various performance metrics, no details are given in terms of the design of the pulse shape to be employed. The authors directly use the standard squared root-raised cosine spectrum pulse. In this paper, we will show that a pulse shape that satisfies the orthogonality conditions for OQAM-OFDM will also satisfy orthogonality with WCP-COQAM for a special selection of the phase term in the transmitted signal of WCP-COQAM. Therefore, efficient and orthogonal pulse shape design algorithms that are proposed for OQAM-OFDM \cite{Bolcskeidzak} can also be used for WCP-COQAM.

The organization of the paper will be as follows. First, the signal models for OQAM-OFDM and WCP-COQAM will be given in Section \ref{sec:signal_model}. In Section \ref{sec:orth_cond}, the orthogonality conditions for OQAM-OFDM and WCP-COQAM will be stated. The proof that a pulse satisfying orthogonality with OQAM-OFDM will also satisfy orthogonality with WCP-COQAM is presented in Appendix \ref{sec:proof_eqv_OQAM_OFDM_WCP_COQAM}. Section \ref{sec:sim_res} verifies the proof with simulation based results. Finally, Section \ref{sec:conclusion} concludes the paper.

\section{Signal Model}
\label{sec:signal_model}
The transmitted signal with OQAM-OFDM can be written as \cite{Bolcskeidzak}
\begin{equation}
\label{eqn:OQAM_transmitted_signal}
\begin{split}
x_{OOM}[n]&=\sum_{k=0}^{K-1}\bigg[\sum_{m={-\infty}}^\infty d_{k,m}^\Re p[n-mK] \\
           &+\sum_{m=-\infty}^\infty jd_{k,m}^\Im p[n+K/2-mK]\bigg]e^{j2\pi\frac{k}{K}(n-\alpha/2)},
\end{split}
\end{equation}
where $d_{k,m}$ is the transmitted data symbol at the $k^{th}$ subcarrier and at the $m^{th}$ symbol timing interval, $d_{k,m}^\Re$ is the real part and $d_{k,m}^\Im$ is the imaginary part of the transmitted symbol $d_{k,m}$, $K$ is equal to the total number of subcarriers and $\alpha=K/2-1$ to be able to design a prototype filter $p[n]$ of length $MK$ that creates no ISI and ICI \cite{Bolcskeidzak}. The filter length should be $MK$ since it will be potentially used with WCP-COQAM. Moreover, the transmitted signal with WCP-COQAM is \cite{lin2014multi}
\begin{equation}
\label{eqn:WCP-COQAM transmitted waveform}
\begin{split}
\hfill x&_{WCP-COQAM}[n]= \\
&\sum_{k=0}^{K-1}\sum_{m={0}}^{2M-1} \tilde{d}_{k,m}p[(n-mK/2)_N]e^{j2\pi\frac{k}{K}(n-D/2)}e^{j\phi_{k,m}}
\end{split}
\end{equation}
where $M$ is the number of time slots in a WCP-COQAM frame, $p[(n-mK)_N]$ corresponds to circular shift by $mK$ with modulo $N=MK$, $\tilde{d}_{k,m}$ are real valued data symbols and $D=MK-1$ \cite{lin2014multi}. Note that a GFDM or WCP-COQAM frame is a block that consists of multiple time slots and a single cyclic prefix. Such a frame structure is explained in detail in \cite{GFDM_5G}. Moreover, since it has been stated in \cite{Siohan2002_May} that there is not a unique selection for the phase term $\phi_{k,m}$ for OQAM modulation schemes, we prefer to choose that $\phi_{k,m}=e^{j\pi k(M-1/2)}M(m)$, where $M(m)=1$ if $m$ is even and $M(m)=j$ if $m$ is odd. Such a selection of $\phi_{k,m}$ term will yield a transmitted signal for WCP-COQAM that has a similar form as the transmitted signal for OQAM-OFDM given in (\ref{eqn:OQAM_transmitted_signal}). In that case, we will be able to relate the orthogonality conditions for OQAM-OFDM written according to the signal model in (\ref{eqn:OQAM_transmitted_signal}) to the orthogonality conditions for WCP-COQAM. For $\phi_{k,m}=e^{j\pi k(M-1/2)}M(m)$, defining $m'=m/2$ and $d_{k,m'}^\Re=\tilde{d}_{k,m}$ when $m$ is even and $m''=(m-1)/2$ and $d_{k,m''}^\Im=\tilde{d}_{k,m}$ when $m$ is odd, the transmitted WCP-COQAM signal can also be expressed as
\begin{equation}
\label{eqn:WCP-COQAM transmitted waveform2}
\begin{split}
x_{WCP-COQAM}&[n]=\sum_{k=0}^{K-1}\bigg[\sum_{m={0}}^{M-1} d_{k,m}^\Re g_k[n]p[(n-mK)_N] \\
           &+\sum_{m=0}^{M-1} jd_{k,m}^\Im g_k[n]p[(n+K/2-mK)_N]\bigg],
\end{split}
\end{equation}
where $g_k[n]=e^{j2\pi\frac{k}{K}(n-\alpha/2)}$ and $\alpha=K/2-1$. Equivalence of (\ref{eqn:WCP-COQAM transmitted waveform}) and (\ref{eqn:WCP-COQAM transmitted waveform2}) when $\phi_{k,m}=e^{j\pi k(M-1/2)}M(m)$ is proved in Appendix \ref{sec:proof_twosignal_models_WCP}.
\section{Orthogonality Conditions}
\label{sec:orth_cond}
When MF receiver is used, the orthogonality conditions for OQAM-OFDM are \cite{Bolcskeidzak}
\begin{equation}
\label{eqn:OQAM_OFDM_orthogonality}
\left.\Re\left\{ p[n-mK]e^{j2\pi\frac{v}{K}(n-\alpha/2)}\right\}\Asterisk\tilde{p}[n] \right|_{n=0} =\delta[m]\delta[v],
\end{equation}
\begin{equation}
\begin{split}
\label{eqn:OQAM_OFDM_orthogonality2}
\Im\Big\{jp[n+K/2-mK]e^{j2\pi\frac{v}{K}(n-\alpha/2)}\Big\}\Asterisk\tilde{p}&[n-K/2] \Big|_{n=0} \\ 
&=\delta[m]\delta[v],
\end{split}
\end{equation}
\begin{equation}
\label{eqn:OQAM_OFDM_orthogonality3}
\left.\Re\left\{jp[n+K/2-mK]e^{j2\pi\frac{v}{K}(n-\alpha/2)}\right\}\Asterisk\tilde{p}[n] \right|_{n=0} =0,
\end{equation}
\begin{equation}
\label{eqn:OQAM_OFDM_orthogonality4}
\left.\Im\left\{p[n-mK]e^{j2\pi\frac{v}{K}(n-\alpha/2)}\right\}\Asterisk\tilde{p}[n-K/2] \right|_{n=0} =0.
\end{equation}
where $\Asterisk$ is the linear convolution operator, $\delta[m]=1$ if $m=0$ and $\delta[m]=0$, otherwise. Moreover, when MF receiver is used the orthogonality conditions for WCP-COQAM are
\begin{equation}
\label{eqn:WCP-COQAM_orthogonality1}
\sum_{n=0}^{N-1}\Re\left\{{p[(n-mK)_N]e^{j2\pi\frac{v}{K}(n-\alpha/2)}p^*[n]}\right\}=\delta[(m)_M]\delta[v],
\end{equation}
\begin{equation}
\begin{split}
\label{eqn:WCP-COQAM_orthogonality2}
\sum_{n=0}^{N-1}\Re\Big\{p[(n+K/2-mK)_N]e^{j2\pi\frac{v}{K}(n-\alpha/2)}&p^*[(n+K/2)_N]\Big\}\\
&=\delta[(m)_M]\delta[v],
\end{split}
\end{equation}
\begin{equation}
\label{eqn:WCP-COQAM_orthogonality3}
\sum_{n=0}^{N-1}\Re\left\{jp[(n+K/2-mK)_N]e^{j2\pi\frac{v}{K}(n-\alpha/2)}p^*[n]\right\}=0,
\end{equation}
\begin{equation}
\label{eqn:WCP-COQAM_orthogonality4}
\sum_{n=0}^{N-1}\Re\left\{p[(n-mK)_N]e^{j2\pi\frac{v}{K}(n-\alpha/2)}jp^*[(n+K/2)_N]\right\}=0,
\end{equation}
where $\delta[(m)_M]$ is defined as
\begin{equation}
\begin{cases}
\delta[(m)_M]=1, & \text{if } m=LM \\
0,				 & \text{otherwise},
\end{cases}
\end{equation}
where $L$ is an integer. Considering the orthogonality conditions given in (\ref{eqn:OQAM_OFDM_orthogonality})-(\ref{eqn:OQAM_OFDM_orthogonality4}) and (\ref{eqn:WCP-COQAM_orthogonality1})-(\ref{eqn:WCP-COQAM_orthogonality4}), the proof that if a pulse with an impulse response of length $MK$ satisfies the orthogonality conditions with OQAM-OFDM, it will also satisfy the orthogonality conditions for WCP-COQAM is presented in Appendix \ref{sec:proof_eqv_OQAM_OFDM_WCP_COQAM}. This proof is considered to be necessary, although in \cite{TS_OQAM_GFDM}, it has been stated that pulse shaping filters that are designed for OQAM-OFDM will directly satisfy orthogonality with WCP-COQAM without such a proof.
Furthermore, the proof will also be verified by simulation based results. A Gaussian and a raised cosine spectrum (RC) pulse, which does not have orthogonality, will be orthogonalized with the discrete Zak transform (DZT) based algorithm defined for OQAM-OFDM in \cite{Bolcskeidzak}. The non-orthogonal Gaussian pulse can be defined as
\begin{equation}
\label{Gaussian_pulse}
p_{Gauss}[n]=\frac{\sqrt{\pi}}{\beta}\exp^{-\frac{\pi^2n}{\beta^2N}},
\end{equation}
where $\beta$ is related to its variance in time domain. The orthogonalized version of the Gaussian pulse or RC pulse will be used in error rate performance simulations. The expectation is to observe the same error rate performance for WCP-COQAM with orthogonalized Gaussian or RC pulses as OFDM, if WCP-COQAM really has orthogonality.
\section{Simulation Results}
\label{sec:sim_res}
\begin{figure}[htbp]
   \centering
\includegraphics[width=0.8\columnwidth]{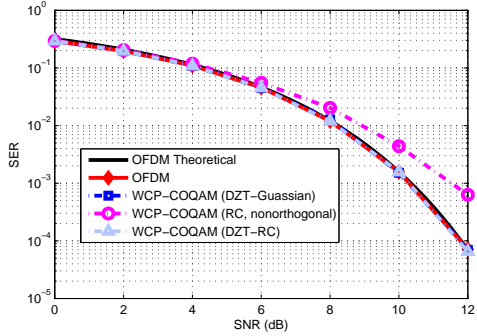}
\caption{SER v.s SNR for OFDM and WCP-COQAM with the orthogonalized RC with roll-off=0.3 or Gaussian pulse with $\beta=0.1$}
\label{fig:SER_SNR_Gaussian_AWGN_alpha0.1}
\end{figure}

The simulation parameters are selected as follows. The number of subcarriers and the number of WCP-COQAM symbols in a WCP-COQAM frame are selected to be 128 and 9, respectively. Constellation order is 4 (QPSK) and the channel type is additive white Gaussian noise (AWGN) channel. A Gaussian pulse with $\beta=0.1$ or an RC pulse with roll-off 0.3 is orthogonalized with the DZT based algorithm to be used with WCP-COQAM. Monte-Carlo trials continue until 100 frame errors are collected. With these parameters, error rate performances are obtained as in Figure~\ref{fig:SER_SNR_Gaussian_AWGN_alpha0.1}.

To illustrate the performance with a nonorthogonal pulse, the error rate performance of WCP-COQAM with RC pulse is included. As can be observed from Figure~\ref{fig:SER_SNR_Gaussian_AWGN_alpha0.1} there is a performance loss when non-orthogonal RC pulse is used. The curve with solid black line indicates the approximate theoretical AWGN performance of OFDM, whose closed form expression is borrowed from \cite{bahai2004multi}. As expected, the OFDM error rate performance obtained with Monte-Carlo based simulations are the same as the theoretical error rate curve. Moreover, it can be observed that the error rate performance of WCP-COQAM with orthogonalized Guassian or RC pulse with DZT based algorithm (corresponding curves are named as DZT-Gaussian or DZT-RC) gives the same performance as OFDM. This verifies that the DZT based algorithm defined for OQAM-OFDM also yields orthogonal pulses for WCP-COQAM.
\section{Conclusion}
\label{sec:conclusion}
In this paper, we prove that a pulse satisfying the orthogonality conditions defined for OQAM-OFDM also satisfies orthogonality when it is used with WCP-CQOAM for a special selection of $\phi_{k,m}$ term in the WCP-COQAM transmitted signal. Therefore, computationally efficient pulse-shape design methods based on DZT developed in \cite{Bolcskeidzak} can also be used with WCP-COQAM. In order to verify this argument, the error rate performances of OFDM and WCP-COQAM are compared when an orthogonalized Gaussian or RC pulse with DZT based algorithm is employed in WCP-COQAM. The simulations yielded similar performance for the two modulation types, which verifies that orthogonality is satisfied in WCP-COQAM.

\appendices
\section{}
\label{sec:proof_twosignal_models_WCP}
\textbf{Theorem:} (\ref{eqn:WCP-COQAM transmitted waveform}) and (\ref{eqn:WCP-COQAM transmitted waveform2}) are equivalent when the phase term in (\ref{eqn:WCP-COQAM transmitted waveform}) $\phi_{k,m}=e^{j\pi k(M-1/2)}M(m)$.

\textbf{Proof:} Replacing $\phi_{k,m}=e^{j\pi k(M-1/2)}M(m)$ and $D=MK-1$ in (\ref{eqn:WCP-COQAM transmitted waveform}) one can get \vspace{-5pt}
\begin{equation}
\label{eqn:WCP-COQAM transmitted waveform6}
\begin{split}
\hfill x&_{WCP-COQAM}[n] \\
&=\sum_{k=0}^{K-1}\sum_{m={0}}^{2M-1} \tilde{d}_{k,m}p[(n-mK/2)_N]e^{j2\pi\frac{k}{K}(n-\frac{K/2-1}{2})}M(m).
\end{split}
\end{equation}
Rewriting the summation term in (\ref{eqn:WCP-COQAM transmitted waveform6}) with index $m$ as a summation of two summations, \vspace{-5pt}
\begin{equation}
\label{eqn:WCP-COQAM transmitted waveform7}
\begin{split}
x&_{WCP-COQAM}[n]\\
=&\sum_{k=0}^{K-1}\bigg\{\sum_{m_1\in M_1} \tilde{d}_{k,m_1} g_k[n]p[(n-m_1K/2)_N]M(m_1) \\
           &\hspace{5pt}+\hspace{12pt}\sum_{m_2\in M_2} \tilde{d}_{k,m_2} g_k[n]p[(n-m_2K/2)_N]M(m_2)\bigg\},
\end{split}
\end{equation}
where $M_1=\{0, 2, 4, ..., 2M-2\}$, $M_2=\{1, 3, 5, ..., 2M-1\}$ and $g_k[n]=e^{j2\pi\frac{k}{K}(n-\frac{K/2-1}{2})}$. Defining $d_{k,m_1'}^\Re=\tilde{d}_{k,m_1}$ $\forall k$, where $m_1'=m_1/2$ and $d_{k,m_2'}^\Im=\tilde{d}_{k,m_2}$ $\forall k$, where $m_2'=(m_2+1)/2$, (\ref{eqn:WCP-COQAM transmitted waveform7}) can also be expressed as \vspace{-5pt}
\begin{equation}
\label{eqn:WCP-COQAM transmitted waveform8}
\begin{split}
x&_{WCP-COQAM}[n]\\
=&\sum_{k=0}^{K-1}\bigg\{\sum_{m_1'=0}^{M-1} d_{k,m_1'}^\Re g_k[n]p[(n-m_1'K)_N] \\
           &\hspace{5pt}+\hspace{12pt}\sum_{m_2'=1}^{M} jd_{k,m_2'}^\Im g_k[n]p[(n+K/2-m_2'K)_N]\bigg\},
\end{split}
\end{equation}
since $M(m)=1$ if $m$ is even and $M(m)=j$ if $m$ is odd. Moreover, since $p[(n+K/2-m_2'K)_N]\big|_{m_2'=M}=p[(n+K/2-m_2'K)_N]\big|_{m_2'=0}$ and defining $d_{k,0}^\Im=d_{k,M}^\Im$ $\forall k$, the limits for $m_2'$ can be written from $m_2'=0$ to $m_2'=M-1$ in (\ref{eqn:WCP-COQAM transmitted waveform8}). In this case (\ref{eqn:WCP-COQAM transmitted waveform8}) has the same form as in (\ref{eqn:WCP-COQAM transmitted waveform2}) which concludes the proof.
\section{}
\label{sec:proof_eqv_OQAM_OFDM_WCP_COQAM}
\textbf{Theorem:} If a pulse shape $p[n]$ satisfies the orthogonality conditions defined for OQAM-OFDM, which are (\ref{eqn:OQAM_OFDM_orthogonality})-(\ref{eqn:OQAM_OFDM_orthogonality4}), it will also satisfy the orthogonality conditions for WCP-COQAM, given in (\ref{eqn:WCP-COQAM_orthogonality1})-(\ref{eqn:WCP-COQAM_orthogonality4}).

\textbf{Proof:} Define $s_{\beta,\gamma}[m,v]$ such that \vspace{-5pt}
\begin{equation}
\label{eqn:s[m]}
s_{\beta,\gamma}[m,v]=\sum_{n=0}^{N-1}p[(n-mK+\beta)_N]g_v[n]p^*[(n+\gamma)_N]
\end{equation}
where $g_v[n]=e^{j2\pi\frac{v}{K}(n-\alpha/2)}$. Assume that $m\in\{1,2,\cdots,M\}$. Moreover, also assume that $\beta$ is either equal to zero or $K/2$. Towards the end of the proof, the reason for such assumptions will be clear. With these assumptions, $-MK\leq\beta-mK \leq 0$. In this case, the term $p[(n-mK+\beta)_N]$ in (\ref{eqn:s[m]}) can also be expressed as follows.
\begin{equation}
\begin{split}
\label{eqn:p[n]_decomposed_forward}
&p[(n-mK+\beta)_N] \\
&=\begin{cases}
p[n+(M-m)K+\beta],&\text{if } 0\leq n \leq mK-\beta-1 \\
p[n-mK+\beta],&\text{if } mK-\beta \leq n \leq MK-1,
\end{cases}
\end{split}
\end{equation}
Moreover, also assume that $\gamma$ is equal to zero or $K/2$. This will bring about $p[(n+\gamma)_N]$ being expressed as follows.
\begin{equation}
\label{eqn:p[n]_decomposed_backward}
\begin{split}
&p[(n+\gamma)_N] \\
&=\begin{cases}
p[n+\gamma], & \text{if  } 0\leq n \leq MK-\gamma-1 \\
p[n-MK+\gamma], & \text{if  } MK-\gamma \leq n \leq MK-1, \\
\end{cases}
\end{split}
\end{equation}
Using (\ref{eqn:p[n]_decomposed_forward}) and (\ref{eqn:p[n]_decomposed_backward}), and also assuming $MK-\gamma-1\geq mK-\beta-1$, (\ref{eqn:s[m]}) can be rewritten as
\begin{equation}
\label{eqn:s[m]_sum}
\begin{split}
s_{\beta,\gamma}[m,v]&=\hspace{-10pt}\sum_{n=0}^{mK-\beta-1}p[n+(M-m)K+\beta]g_v[n]p^*[n+\gamma] \\
	 &+\hspace{-10pt}\sum_{n=mK-\beta}^{MK-\gamma-1} p[n-mK+\beta]g_v[n]p^*[n+\gamma] \\
     &+\hspace{-10pt}\sum_{n=MK-\gamma}^{MK-1}p[n-mK+\beta]g_v[n]p^*[n-MK+\gamma].
\end{split}
\end{equation}
Changing the summation limits for the third summation term in (\ref{eqn:s[m]_sum}), one can get
\begin{equation}
\label{eqn:s[m]_sum2}
\begin{split}
s_{\beta,\gamma}[m,v]&=\hspace{-10pt}\sum_{n=0}^{mK-\beta-1}p[n+(M-m)K+\beta]g_v[n]p^*[n+\gamma] \\
	 &+\hspace{-10pt}\sum_{n=mK-\beta}^{MK-\gamma-1}p[n-mK+\beta]g_v[n]p^*[n+\gamma] \\
     &+\hspace{-4pt}\sum_{n=-\gamma}^{-1}p[n+(M-m)K+\beta]g_v[n]p^*[n+\gamma].
\end{split}
\end{equation}
since $g_v[n]=g_v[n+MK]$. Note that in (\ref{eqn:s[m]_sum2}), the first and the third summation terms can be combined to obtain
\begin{equation}
\label{eqn:s[m]_sum3}
\begin{split}
s_{\beta,\gamma}[m,v]&=\sum_{n=-\gamma}^{mK-\beta-1}p[n+(M-m)K+\beta]g_v[n]p^*[n+\gamma] \\
 &+ \sum_{n=mK-\beta}^{MK-\gamma-1}p[n-mK+\beta]g_v[n]p^*[n+\gamma].
\end{split}
\end{equation}
Moreover, the two summation terms in (\ref{eqn:s[m]_sum3}) can also be written as follows.
\begin{equation}
\label{eqn:convolution_relations}
\begin{aligned}
\sum_{n=-\gamma}^{mK-\beta-1}&p[n+(M-m)K+\beta]g_v[n]p^*[n+\gamma] \\
&= p[n+(M-m)K+\beta]g_v[n]\Asterisk\tilde{p}[n-\gamma]\big\rvert_{n=0} \\
\sum_{n=mK-\beta}^{MK-\gamma-1}&p[n-mK+\beta]g_v[n]p^*[n+\gamma] \\
&=p[n-mK+\beta]g_v[n]\Asterisk\tilde{p}[n-\gamma]\big|_{n=0}, 
\end{aligned}
\end{equation}
where $\Asterisk$ is the convolution operator and $\tilde{p}[n]=p^*[-n]$. Using (\ref{eqn:s[m]_sum3}) and (\ref{eqn:convolution_relations}), (\ref{eqn:s[m]}) can also be written as
\begin{equation}
\label{eqn:s[m]_reexpressed}
\begin{split}
s_{\beta,\gamma}[m,v]&=p[n+(M-m)K+\beta]g_v[n]\Asterisk\tilde{p}[n-\gamma]\big\rvert_{n=0} \\
&+p[n-mK+\beta]g_v[n]\Asterisk\tilde{p}[n-\gamma]\big|_{n=0}.
\end{split}
\end{equation}
Moreover, the orthogonality conditions for WCP-COQAM, which are (\ref{eqn:WCP-COQAM_orthogonality1})-(\ref{eqn:WCP-COQAM_orthogonality4}), can be written in terms of $s_{\beta,\gamma}[m,v]$ as follows.
\begin{equation}
\label{eqn:WCP-COQAM_orthogonality1_intmsof_s}
\Re\left\{s_{\beta=0,\gamma=0}[m,v]\right\}=\delta[(m)_M]\delta[v],
\end{equation}
\begin{equation}
\label{eqn:WCP-COQAM_orthogonality2_intmsof_s}
\Im\left\{js_{\beta=K/2,\gamma=K/2}[m,v]\right\}=\delta[(m)_M]\delta[v],
\end{equation}
\begin{equation}
\label{eqn:WCP-COQAM_orthogonality3_intmsof_s}
\Re\left\{js_{\beta=K/2,\gamma=0}[m,v]\right\}=0,
\end{equation}
\begin{equation}
\label{eqn:WCP-COQAM_orthogonality4_intmsof_s}
\Im\left\{s_{\beta=0,\gamma=K/2}[m,v]\right\}=0.
\end{equation}
Replacing $s_{\beta,\gamma}$ in (\ref{eqn:WCP-COQAM_orthogonality1_intmsof_s})-(\ref{eqn:WCP-COQAM_orthogonality4_intmsof_s}) using (\ref{eqn:s[m]_reexpressed}), and since $g_v[n]=e^{j2\pi\frac{v}{K}(n-\alpha/2)}$, one can get
\begin{equation}
\label{eqn:WCP-COQAM_orthogonality1_final}
\begin{split}
&\Re\{p[n+(M-m)K]e^{j2\pi\frac{v}{K}(n-\alpha/2)}\Asterisk\tilde{p}[n]\big|_{n=0}\} \\
&+\Re\{p[n-mK]e^{j2\pi\frac{v}{K}(n-\alpha/2)}\Asterisk\tilde{p}[n]\big|_{n=0}\}=\delta[(m)_M]\delta[v],
\end{split}
\end{equation}
\begin{equation}
\nonumber
\Im\{jp[n+(M-m)K+K/2] e^{j2\pi\frac{v}{K}(n-\alpha/2)}\Asterisk\tilde{p}[n-K/2]\big|_{n=0}\},
\end{equation}
\begin{equation}
\label{eqn:WCP-COQAM_orthogonality2_final}
\begin{split}
+\Im\{jp[n-mK+K/2]e^{j2\pi\frac{v}{K}(n-\alpha/2)}\Asterisk\tilde{p}&[n-K/2]\big|_{n=0}\}\\
&=\delta[(m)_M]\delta[v],
\end{split}
\end{equation}\begin{equation}
\label{eqn:WCP-COQAM_orthogonality3_final}
\begin{split}
&\Re\{jp[n+(M-m)K+K/2]e^{j2\pi\frac{v}{K}(n-\alpha/2)}\Asterisk\tilde{p}[n]\big|_{n=0}\} \\
&+\Re\{jp[n-mK+K/2]e^{j2\pi\frac{v}{K}(n-\alpha/2)}\Asterisk\tilde{p}[n]\big|_{n=0}\}=0,
\end{split}
\end{equation}\begin{equation}
\label{eqn:WCP-COQAM_orthogonality4_final}
\begin{split}
&\Im\{p[n+(M-m)K]e^{j2\pi\frac{v}{K}(n-\alpha/2)}\Asterisk\tilde{p}[n-K/2]\big|_{n=0}\} \\
&+\Im\{p[n-mK]e^{j2\pi\frac{v}{K}(n-\alpha/2)}\Asterisk\tilde{p}[n-K/2]\big|_{n=0}\}=0.
\end{split}
\end{equation}
Consider the case $m\in\{1,2,\cdots,M-1,M\}$. For these values of $m$, the assumed conditions  $MK-\gamma-1\geq mK-\beta-1$ and $-MK \leq \beta-mK \leq 0$ in the proof hold when $\beta=\gamma=0$ or $\beta=\gamma=K/2$ or $(\beta,\gamma)=(K/2,0)$, which corresponds to the cases in (\ref{eqn:WCP-COQAM_orthogonality1_intmsof_s}), (\ref{eqn:WCP-COQAM_orthogonality2_intmsof_s}) and (\ref{eqn:WCP-COQAM_orthogonality3_intmsof_s}). Therefore, it is valid that (\ref{eqn:WCP-COQAM_orthogonality1_intmsof_s}), (\ref{eqn:WCP-COQAM_orthogonality2_intmsof_s}), (\ref{eqn:WCP-COQAM_orthogonality3_intmsof_s}) are equivalent to (\ref{eqn:WCP-COQAM_orthogonality1_final}), (\ref{eqn:WCP-COQAM_orthogonality2_final}), (\ref{eqn:WCP-COQAM_orthogonality3_final}), respectively. For these values of $m$, (\ref{eqn:OQAM_OFDM_orthogonality}) implies (\ref{eqn:WCP-COQAM_orthogonality1_final}) when $p[n]$ is real valued. Similarly, for the same $m$ values, when $p[n]$ is real valued, (\ref{eqn:WCP-COQAM_orthogonality2_final}) and (\ref{eqn:WCP-COQAM_orthogonality3_final}) are also satisfied by (\ref{eqn:OQAM_OFDM_orthogonality2}) and (\ref{eqn:OQAM_OFDM_orthogonality3}), respectively. Note also that, if the orthogonality conditions in (\ref{eqn:WCP-COQAM_orthogonality1_intmsof_s})-(\ref{eqn:WCP-COQAM_orthogonality3_intmsof_s}) hold for $m=1,2,\cdots,M-1,M$, they will also hold for any possible value of $m$, since $s_{\beta,\gamma}[m,v]=s_{\beta,\gamma}[m+PM,v]$ for any integer $P$ value. 

The remaining case in the proof is the satisfaction of (\ref{eqn:WCP-COQAM_orthogonality4_intmsof_s}). In (\ref{eqn:WCP-COQAM_orthogonality4_intmsof_s}), note that $\beta=0,\gamma=K/2$. For these values, choose another set for the possible values of $m$ as $\{{0,1,2,\cdots,M-1}\}$. For these values of $m$, $\beta$, $\gamma$, the assumed conditions in the proof $MK-\gamma-1\geq mK-\beta-1$ and $-MK \leq \beta-mK \leq 0$ hold. Therefore, (\ref{eqn:WCP-COQAM_orthogonality4_intmsof_s}) is equivalent to (\ref{eqn:WCP-COQAM_orthogonality4_final}). Since for the specified set of $m$ values (\ref{eqn:WCP-COQAM_orthogonality4_final}) is also satisfied by (\ref{eqn:OQAM_OFDM_orthogonality4}) when $p[n]$ is real, the proof concludes. For the other possible values of $m$, (\ref{eqn:WCP-COQAM_orthogonality4_intmsof_s}) will also hold since $s_{\beta,\gamma}[m,v]=s_{\beta,\gamma}[m+PM,v]$ for any integer $P$ value.

% Can use something like this to put references on a page
% by themselves when using endfloat and the captionsoff option.
\ifCLASSOPTIONcaptionsoff
  \newpage
\fi

\end{document}